\documentclass[aps,prl,twocolumn,a4paper,10pt]{revtex4-1}

\usepackage[T1]{fontenc}		
\usepackage[english]{babel}		
\usepackage[latin1]{inputenc}	
\usepackage{times}

\usepackage{amsmath}			
\usepackage{amssymb}			
\usepackage{bbm}				
\usepackage{mathtools}
\usepackage{simpler-wick}
\usetikzlibrary{tikzmark}
\DeclarePairedDelimiterX\Dirbraket[3]{\langle}{\rangle}%
{#1\,\delimsize\vert\,\mathopen{}#2\,\delimsize\vert\,\mathopen{}#3}

\usepackage[colorlinks=true, a4paper=true, pdfstartview=FitV,linkcolor=blue, citecolor=blue, urlcolor=blue]{hyperref}

\usepackage{graphicx}			
\usepackage{graphics}			
\DeclareGraphicsExtensions{.pdf}
\usepackage{color}		

\newcommand{\bea}{\begin{eqnarray}}
\newcommand{\eea}{\end{eqnarray}}

\newcommand{\bk}{\mathbf{k}}

\begin{document}

\title{Spectral shift technique for strongly correlated lattice fermions}

\author{Johan~Carlstr\"om }
\affiliation{Department of Physics, Stockholm University, 106 91 Stockholm, Sweden}
\date{\today}

\begin{abstract}
We present a development of strong-coupling diagrammatic techniques which relies on integrating out mean-field-like paths prior to conducting the expansion. This makes it possible to expand around a state with a quasiparticle spectrum that takes into account all kinetic effects that do not explicitly depend on nonlocal correlations. These paths contribute most of the kinetic energy in correlated systems, and so this protocol provides a starting point for the expansion that more closely resembles the full theory. Comparisons to existing exact results clearly confirm this. 
\end{abstract}
\maketitle


\section{Introduction}
Accurate theoretical treatment of fermionic many-body physics remains one of the most challenging problems in condensed matter physics due to the sign problem, which prevents conventional Monte Carlo methods from being effective. 
This has spurred an extensive development of computational methods that seek to overcome this fundamental problem. While very significant progress has been made, these techniques are still limited in either accuracy or applicability. 

In strongly correlated systems, key advances have been made with methods like DMFT \cite{PhysRevLett.62.324}, with extensions based on diagrams techniques \cite{PhysRevB.75.045118,PhysRevB.77.033101}, cluster generalizations \cite{PhysRevLett.106.047004,PhysRevLett.110.216405,RevModPhys.77.1027}, and similar ideas  \cite{PhysRevLett.112.196402,PhysRevB.92.115109}. Other important examples include DMRG \cite{PhysRevLett.69.2863}, wave function methods \cite{PhysRevLett.87.217002,PhysRevB.95.024506} and auxiliary-field quantum Monte Carlo \cite{PhysRevB.39.839,PhysRevLett.62.591,PhysRevB.40.506,afqmc}. 
Many of these techniques have produced clearly relevant solutions for strongly correlated electronic systems, most notably the cuprate superconductors. Key examples of this include anti-ferromagnetism, striped states, pseudo-gap physics, and d-wave superconductivity. 
While the results clearly show that current numerical techniques can reproduce key phases of correlated systems, these protocols continue to display notable discrepancies. These appear both when comparing different techniques but also when changing details of the implementation, like the discretization \cite{2006cond.mat.10710S}. The sensitivity that these systems show with respect to computational protocols may be rooted in competition between phases that are closely situated in terms of free energy \cite{2006cond.mat.10710S,Dagotto257}. Despite significant progress recently, key parameter regions of the Hubbard model remain poorly understood \cite{PhysRevX.5.041041}. 

While the aforementioned techniques generally depend on an approximative treatment of many-body theories, a distinct category of unbiased numerical protocols is now evolving rapidly. These methods generally depend on a series expansion such that the only systematic error is due to the truncation of this series. Thus, the result is asymptotically exact, and convergence can be established by examining how an observable changes with the expansion order. If the series is convergent and can be computed to a sufficiently high order, then observables can be obtained with known and very small error bars. 

Numerical linked cluster expansion \cite{PhysRevLett.97.187202} (NLCE)  is based on exact diagonalization of small embedded clusters. Convergence to the macroscopic result can be observed with increasing cluster size. This technique is applicable to spin models \cite{PhysRevE.75.061118} and strongly correlated itinerant fermions \cite{PhysRevE.75.061119,PhysRevE.89.063301,PhysRevA.84.053611}. The main limitation of this method is that controllable results are only obtained at relatively high temperatures. 

Diagrammatic Monte Carlo \cite{Van_Houcke_2010} is based on the stochastic sampling of the diagrammatic expansion and is also a controllable technique since convergence can be observed with respect to expansion order. With proper resummation techniques, results can also be extracted from a divergent series \cite{PhysRevLett.121.130405}.  This method has proven to be applicable to a very wide range of systems, including the unitary fermi gas \cite{DiagramVsEmulator,2018unitary}, frustrated spins \cite{2020SpinIce,PhysRevLett.116.177203}, graphene \cite{PhysRevLett.118.026403}, topological semimetals \cite{PhysRevB.98.241102,2018Symmetry} and lattice fermions \cite{Simkovic2021weak}--including the pairing symmetry of the Hubbard model in the fermi liquid regime \cite{0295-5075-110-5-57001,PhysRevB.104.L020507}. 

In the strongly correlated regime, diagrammatic techniques have typically struggled with a large expansion parameter that often results in a non-analytic behavior of the series. This has motivated an extensive effort to overcome this problem, either by changing the analytical structure of the expansion or even mapping the problem onto an entirely new description. 

Shifted/homotopic action operates on the principle of altering the starting point of the expansion, and thus also the expansion parameter so as to place the parameter range of interest within the radius of convergence \cite{PhysRevB.93.161102,kim2020homotopic}. By extracting the analytical structure of the self-energy, it becomes possible to also reconstruct it in the non-perturbative region \cite{PhysRevB.100.121102}. With efficient sampling protocols \cite{PhysRevLett.119.045701,Rossi_2017}, this has thus far provided access to systems with an interaction strength up to $U/t\sim7$. While this is a substantial improvement over conventional diagrammatic Monte Carlo, it still falls short of $U/t\sim 12$, which is relevant for the cuprate superconductors, for example. 

Second fermionization relies on a radical reformulation of the problem by projecting out doubly occupied sites and replacing them with bosons that are subsequently fermionized \cite{PhysRevB.84.073102}. Combined with spin-charge transformation \cite{0953-8984-29-38-385602} and Popov-Fedotov fermionization \cite{JETP.67.535}, this resulted in the first diagrammatic Monte Carlo results well inside the strongly correlated regime \cite{PhysRevB.97.075119}. These were strictly limited to high temperatures due to the complexity of the resulting theory, though the series itself remains convergent also at lower temperatures. In the spin-charge transformed Hubbard model, the hopping integral $t$ is an expansion parameter, and thus, the method is essentially a type of strong-coupling technique. Integrating out the contact interactions and arranging them into effective vertices is the starting point of strong-coupling diagrammatic Monte Carlo (SCDMC) \cite{PhysRevB.103.195147}. This method produces a series that is identical to that of the spin-charge transformed theory but with dramatic improvements in computational complexity that give access to much lower temperatures. 

Here, we present a development of SCDMC that relies on integrating out certain classes of kinetic processes that are mean-field-like. As a result, the state around which we expand is shifted from the atomic theory with peaks at energies $-\mu$ and $U-2\mu$, to a phase with a quasi-particle spectrum that accounts for these contributions. 
The expansion is then only conducted in processes with an explicit dependence on the inter-site correlations of the system. 
We present mean-field solutions for the equation of state, which correspond to order zero in this expansion, and compare these to NLCE and SCDMC. This clearly shows that almost the entire correction to the carrier density form delocalization is provided from these mean-field-like terms. 

\section{Connected vertices for itinerant fermions}
For completeness we review the main idea of SCDMC \cite{PhysRevB.103.195147} applied to the Hubbard model:
\bea
H=\sum_{\langle i,j\rangle} t c_{i,\sigma}^\dagger c_{j,\sigma}+\sum_i U n_{i,\uparrow}n_{i,\downarrow}-\mu n_i 
= \hat{t}\!+\!\hat{U}\!+\!\hat{\mu}.\;\;\;
\label{model}
\eea
Here we have introduced a shorthand notation for the nonlocal part, the contact interaction, and the local bilinear part.  
In Feynman type diagrammatics, the expansion is carried out in the non-bilinear part of the Hamiltonian. Here, we instead proceed to derive a set of connected local vertices, which form the basis for an expansion in the nonlocal processes. To this end, we begin by separating the Hamiltonian into two parts:
\bea
H=H_0+H_1,\;H_0=\hat{\mu}, \;H_1= \hat{t}+\hat{U}.
\eea 

Since $H_0$ is bilinear, we can treat the model (\ref{model}) through expansion in $H_1$ according to
\bea
\langle\hat{M}\rangle\!= \!
\frac{1}{Z}\sum_n \!\frac{(- 1 )^n}{n!} \! \! \int_0^\beta \! \! d \tau_i  
\text{Tr}\{e^{\! \!-\beta \! H_0  }  T \![H_1(\tau_1 ) ...  H_1(\tau_n )  \hat{M} ] \}.\;\;\;\;\label{expansion}
\eea
We may also note that the nonlocal terms are contained in $H_1$, and correspondingly, the bare Greens function is local, so that 
\bea
G^0_{\alpha\beta}(i-j,\tau)=G^0_{\alpha\beta}(\tau)\delta_{i,j}. \label{G0local}
\eea
To simplify the derivation, let us now introduce the following short hand notation for the integral over a time-ordered product: 
\bea
\Gamma_n=\frac{(-1)^n}{n!}\int_0^\beta d\tau_1...d\tau_n T_\tau
\eea
with the generalization
\bea
\Gamma_n\Gamma_m=\frac{(-1)^{n+m}}{n!m!}\int_0^\beta d\tau_1...d\tau_{n+m} T_\tau.
\eea
The expansion in $H_1$ can then be written
\bea
\sum_n\Gamma_n H_1^n=\sum_{n,m} \Gamma_n\Gamma_m\hat{U}^n\hat{t}^m\\
=\sum_{m,n_1,n_2...}\Gamma_m\Gamma_{n_1}... \hat{U}_1^{n_1}\hat{U}_2^{n_2}...\hat{t}^m .\label{spatialDecomp}
\eea
We then express the nonlocal terms (and the observable $\hat{M}$) according to $\hat{t}=t\hat{o}_{\alpha,i}\hat{o}_{\beta,j}$, where $\hat{o}_{\gamma,i}$ is a creation/annihilation operator on the site $i$. Denoting the set of such operators on the site $i$ by $\bar{o}_i=o_{1,i}o_{2,i}...$,
the expansion in connected diagram topologies then takes the form 
\bea
\langle\hat{M}\rangle=\Big\langle\sum_n \Gamma_n t^n \sum_{\bar{x}} \prod_i \sum_{n_i} e^{-\beta H_{0,i}} \Gamma_{n_i}U_i^{n_i} \bar{o}_i\Big\rangle_c.\label{expansion2}
\eea
Here, $\bar{x}$ is the spatial degrees of freedom of the nonlocal terms, while the subscript $c$ implies connected topologies. 
Since the bare Greens function is local, it follows that contractions may be carried out on single lattice sites separately. However, a key problem remains: We are interested specifically in connected topologies, and this is a global property of a diagram. To overcome this obstacle, we cannot simply compute the trace on each lattice site Instead, we have to classify the contractions on a given site according to their connectivity. First, we note, that on the site $i$, we can break out all local terms that are not connected to any external line $\hat{t}$, i.e. 
\bea
\sum_{n} \langle \Gamma_{n} U_i^{n} \bar{o}_i\rangle_{\hat{\mu}}=\sum_{n,m} \langle \Gamma_n U_i^n \bar{o}_i\rangle_{\hat{\mu},e}\langle \Gamma_m U^m   \rangle_{\hat{\mu}}.\label{ext}
\eea
Here, the subscript $e$ denotes contractions such that all  operators are connected to at least one external operator from the set $\bar{o}_i$. 

The second factor of (\ref{ext}) is the partition function of $\hat{\mu}+\hat{U}$, and correspondingly it follows that 
\bea
\sum_{n} \langle \Gamma_n U_i^n \bar{o}_i\rangle_{\hat{\mu},e}=\langle  \bar{o}\rangle_{\hat{\mu}+\hat{U}}.\label{expO}
\eea
Here, it should be noted that the expansion (\ref{expO}) is formally not convergent for large values of $U$ \cite{PhysRevLett.114.156402, PhysRevLett.119.056402}. However, this summation can be rendered analytic by second fermionization of the Hubbard model \cite{PhysRevB.103.195147}, and is then exactly solvable.

The terms in the second factor of (\ref{ext}) do not contribute to any connected topologies and may be discarded, leading to an expansion of the form  
\bea
\sum_n \Gamma_n t^n \sum_{\bar{x}}  \Big[\!\prod_i \sum_{n_i} \langle \Gamma_{n_i}U_i^{n_i} \bar{o}_i\rangle_{\mu,e}\Big]_c.\label{expansion3}
\eea

At this stage, the contractions within the brackets $\langle \rangle$ are connected with at least one nonlocal operator. To make further progress, we have to sort the contractions according to their connectivity with respect to external operators $\bar{o}_i$. 

Thus, we define the {\it connected} set of contractions denoted by $\langle...\rangle_{\mu,c}  $ as the subset of $\langle...\rangle_{\mu,e} $ for which all elements of $\bar{o}_i$ are connected by local contractions. By contrast, {\it disconnected} contractions do not satisfy this criterium. 
The connected vertex can be extracted from (\ref{expO}) via a recursive process where all contributions from disconnected terms are subtracted. 

\begin{figure}[!htb]
\includegraphics[width=\linewidth]{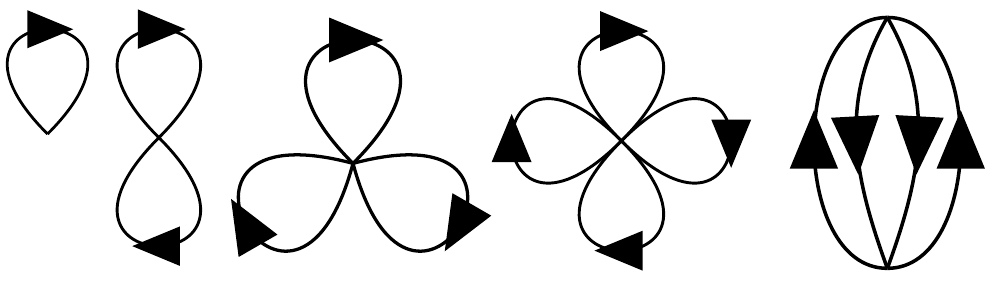}
\caption{
{\bf Strong-coupling expansion.}
At orders up to $N_t=4$, we obtain a total of five skeleton-diagram topologies. Note that the bold hopping defined in Eq. (\ref{bethe}) also has a local part, which allows the lines to close on themselves. The vertices are of the connected type, as defined by Eq. (\ref{recursion}). 
}
\label{hubbardTop}
\end{figure}

\begin{figure}[!htb]
\includegraphics[width=\linewidth]{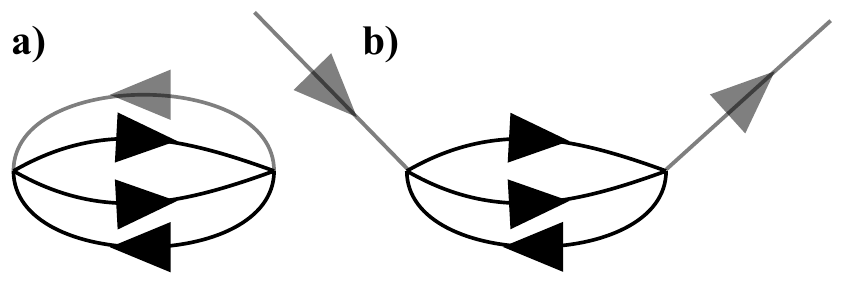}
\caption{
{\bf Measuring lines.} The shaded line in {\bf(a)} is "tagged", and hence represents a fermion entering and leaving the system respectively (b). The reminder of the diagram provides a contribution to the polarization operator of the $t-$line. Using multiple measuring lines, it becomes possible to extract arbitrary correlation functions. 
}
\label{mline}
\end{figure}

The establish the recursive relationship, we begin by sorting the operators $\bar{o}$ according to whether or not they are connected to $o_1\in\bar{o}$ via local contractions. Those that are connected make up the set $A$, while the remaining constitute $\bar{o}\setminus A$. 
For each choice of $A$, we have a set of contractions given by
\bea
\xi_{\bar{o},A}
 \sum_{n,m} \langle \Gamma_n U^n A\rangle_{\hat{\mu},c}\langle \Gamma_m U^m \bar{o}\setminus A\rangle_{\hat{\mu},e},
\eea
where $\xi_{\bar{o},A}$ is a fermionic sign given by
\bea
\xi_{\bar{o},A}=(-1)^c
\eea 
with $c$ denoting the number of fermionic commutations required for the reordering  
\bea
T_\tau \bar{o}\to T_\tau A \times T_\tau(\bar{o} \setminus A).
\eea
To obtain the contribution from all disconnected topologies, we sum over all choices of $A$ that satisfy two criteria: Firstly, $A$ must contain $o_1$ since it is the set of operators connected to it. Secondly, since we are listing disconnected topologies, $\bar{o}\setminus A$ must be a nonempty set, implying that $A$ is a  proper subset of $\bar{o}$. Thus we arrive at the following sum for the contribution from disconnected contractions:
\bea
\sum_{A\subsetneq \bar{o}, \hat{o}_1\in A}
\xi_{\bar{o},A}
 \sum_{n,m} \langle \Gamma_n U^n A\rangle_{\hat{\mu},c}\langle \Gamma_m U^m \bar{o}\setminus A\rangle_{\hat{\mu},e}.\label{disconnected}
\eea
Using (\ref{disconnected}) we can then construct a recursive relation for the connected vertex on the form 
\bea\nonumber
V[\bar{o}]=\sum_n\langle \Gamma_n U^n \bar{o}\rangle_{\hat{\mu},c}=\sum_n\langle \Gamma_n U^n \bar{o}\rangle_{\hat{\mu},e}
\\
\!-\!\sum_{A\subsetneq \bar{o}, \hat{o}_1\in A}\!\xi_{\bar{o},A}
 \sum_{n,m} \langle \Gamma_n U^n A\rangle_{\hat{\mu},c}\langle \Gamma_m U^m \bar{o}\setminus A\rangle_{\hat{\mu},e},\label{recursion}
\eea
Since $\langle \sum_n \Gamma_n U^n \bar{o}\rangle_{\hat{\mu},e}$ can be solved analytically via (\ref{expO}), the recursion (\ref{recursion}) allows the connected vertex to be computed exactly. This is the starting point for strong-coupling diagrammatic Monte Carlo, where connected vertices form the basic building blocks of the series expansion. These vertices are then connected via the nonlocal processes $\hat{t}$ to form connected diagram topologies, see also Fig. \ref{hubbardTop}. Observables are extracted using so-called measuring lines as illustrated in Fig. \ref{mline}: One line is tagged as a measuring line, and represents an incoming/outgoing particle, while the rest of the diagram then provides a contribution to the polarization operator of the hopping-line $t$. Using multiple measuring lines, one can compute arbitrary correlation functions.
 
The dressed hopping line can be obtained via the Bethe-Salpiter equation:
\bea
\tilde{t}(\bk,\omega)=\frac{1}{t^{-1}(\bk)-\Pi_t(\bk,\omega)}.\label{bethe}
\eea
Here, $\Pi_t$ is the polarization operator for the $t$-line, while $\tilde{t}$ represents the dressed hopping. Using eq. (\ref{bethe}), the expansion can be conducted in the dressed hopping $\tilde{t}$ and skeleton diagrams, which contain no insertions form the polarization operator.  
The electron Greens function is obtained form 
 \bea
 G(\bk,\omega)=\frac{1}{\Pi_t^{-1}(\bk,\omega)-t(\bk)}. \label{GreensFunction}
 \eea


\section{The Spectral shift}
A major limitation of strong coupling techniques is that the expansion is conducted around an atomic state, where kinetic corrections to single-particle properties are not taken into account at all but have to be resolved solely by the expansion. 
This means that when computing observables, a substantial part of the effort is spent on determining how delocalization effects renormalize single-particle properties. The basic premise of the spectral shift technique is that strong-coupling techniques can be dramatically improved upon by integrating out certain classes of kinetic processes that are essentially mean-field-like. In doing so, we shift the excitations from the atomic theory with peaks at $-\mu$ and $U-2\mu$ respectively, to a quasiparticle spectrum that reflects the mean-field-like delocalization effects. 

\begin{figure}[!htb]
\includegraphics[width=\linewidth]{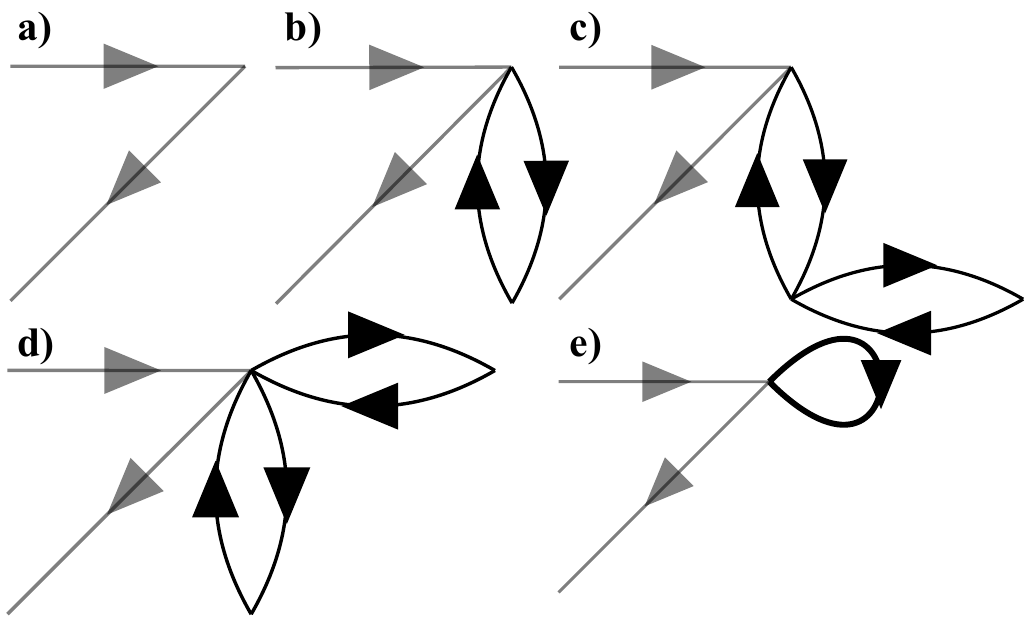}
\caption{
{\bf  Examples of mean-field like diagrams.
}
	The two-legged vertex ({\bf a}) is effectively renormalized by loops of the form ({\bf b}-{\bf d}). These processes do not depend on any correlations in the system, and are in that sense mean-field like. Once the hopping line $t$ is boldified according to Eq. (\ref{bethe}), these terms are represented by self-closing lines ($\tilde{t}$) of the form ({\bf e}).
}
\label{selfTrace}
\end{figure}

The starting point of this technique is the observation that certain classes of terms in the strong-coupling expansion do not depend on inter-site correlations but are completely defined by averages. A few examples of such corrections are given in Fig. \ref{selfTrace}. Notably, when boldifying the hopping line, these contributions are contained in self-closing loops of the form (\ref{selfTrace}, {\bf e}), which depend on the contact-part of $\tilde{t}$. This prompts us to separate the bold hopping line into local and nonlocal parts, respectively, according to
\bea
\tilde{t}(\bk,\omega)=\tilde{L}(\omega)+\tilde{I}(\bk,\omega),
\eea 
where $L$ and $I$ represent local and itinerant processes respectively. 
Denoting the operators due to the local and itinerant parts of $\tilde{t}$ by $\bar{\lambda}$ and $\bar{\iota}$ respectively, we can rewrite  Eq. (\ref{expansion3}) as  
\bea
\sum_n \Gamma_n \sum_{\bar{x}}  \Big[\!\prod_i \sum_{n_i,m_i} \langle \Gamma_{n_i}\tilde{\Gamma}_{m_i}U_i^{n_i} \bar{\lambda}_i^{m_i}\bar{\iota}_i\rangle_{\mu,e}\Big]_c,\label{expansion4}
\eea 
where the subscript $e$ denotes that all terms are connected to at least one element in $\bar{\lambda}$ or $\bar{\iota}$.
Here, we sum over the order of $\tilde{L}$ and $\tilde{I}$ independently. The generalized time-ordered integral is defined as 
\bea
\tilde{\Gamma}_n=\frac{(-1)^n}{n!}\int_0^\beta d\tau_1 d\tau_1'...d\tau_n d\tau_n' T_\tau
\eea
since $\tilde{L}$ is nonlocal in time and thus requires two imaginary time-variables. Using the recursion \ref{recursion}, we can extract vertices 
\bea
V[\bar{\iota},\bar{\lambda}]
\eea
such that all terms in $\bar{\lambda},\bar{\iota}$ are connected by contact interactions $\sim U$. At the next stage, we construct a {\it shifted vertex} as follows: The zero order part of this object is just the ordinary connected vertex. 
\bea
\tilde {V}_0[\bar{\iota}]=V[\bar{\iota}].
\eea 
The first correction takes the form 
\bea
\tilde {V}_1[\bar{\iota}]=\sum_{\{\bar{\lambda}\}} \tilde{\Gamma}_n C(\bar{\lambda})  V[\bar{\iota},\bar{\lambda}],
\eea 
where $n$ is the order in $\tilde{L}$, and $C(\lambda)$ is the set of ways in which the operators $\bar{\lambda}$ can be connected by lines $\tilde{L}(\tau)$, which takes the form of a determinant. 

At second order we obtain
\bea
\tilde {V}_2[\bar{\iota}]=\!\!\!\!\!\!\!\!\!\sum_{A\subset \bar{\iota},\{\bar{\lambda}_a\},\{\bar{\lambda}_b\}}  \!\!\!\!\!\!\!\!\! \tilde{\Gamma}_{n_a}\tilde{\Gamma}_{n_b} C(\bar{\lambda}_a,\bar{\lambda}_b) V[\bar{\iota}\setminus A,\bar{\lambda}_a]V[A,\bar{\lambda}_b].\;\;\;\;
\eea
Here, $C$ is the set of lines $\tilde{L}$ that can be drawn between operators $\bar{\lambda}$ such that the two vertices are connected and form an irreducible topology. The summation over $A$ represents the ways in which the set of operators $\bar{\iota}$ can be split into two sets (note that $A$ is permitted to be empty). 

Thus we can identify a set of rules that apply when calculating the shift: (I) The order of the correction determines the number of connected vertices $V$. (II) The set of nonlocal operators $\bar{\iota}$ may be split in any manner between the vertices. (III) The operators $\bar{\lambda}$ must be connected by lines $\tilde{L}$ such that the vertices are connected and form an irreducible topology. Here, irreducibility implies that a diagram does not contain insertions from the polarization operator $\Pi_t$.

Once the local part of $\tilde{t}$ is integrated out, we can conduct an expansion in shifted vertices $\tilde{V}$, which are then connected by lines that correspond to the nonlocal part of $\tilde{t}$. We refer to this as the {\it shifted expansion}.

\section{Mean-field results}
To test the prelusive conjecture--that corrections to single particle properties from delocalization are to a large extent mean-field-like--we will now establish a form of mean-field theory that is based on an iterative calculation of the two-legged shifted vertex. For the shifted expansion, this corresponds to the order zero. 
At the lowest order, the polarization operator is given by
\bea
\Pi_{t,\sigma}(\tau_1-\tau_2)=\tilde{V}[c^\dagger_{\sigma}(\tau_2)c_{\sigma}(\tau_1)].\label{mft}
\eea
Computing $\Pi_t$ from Eq. (\ref{mft}), we obtain the dressed hopping using the Bethe-Salpiter equation 
\bea
\tilde{t}(\bk,\omega)=\frac{1}{t^{-1}(\bk)-\Pi_t(\omega)},
\eea
where $\Pi_t$ is local and thus momentum-independent. This gives us $\tilde{L}$, which we can use to compute $\tilde{V}_2$. Inserting $\tilde{V}_2$ into Eq. (\ref{mft}) we obtain a new result for $\Pi_t$. Repeating this scheme until convergence, we take into account an infinite class of mean-field-like processes. Single-particle properties can then be extracted from the Greens function using Eq. (\ref{GreensFunction}).

 \begin{figure}[!htb]
\includegraphics[width=\linewidth]{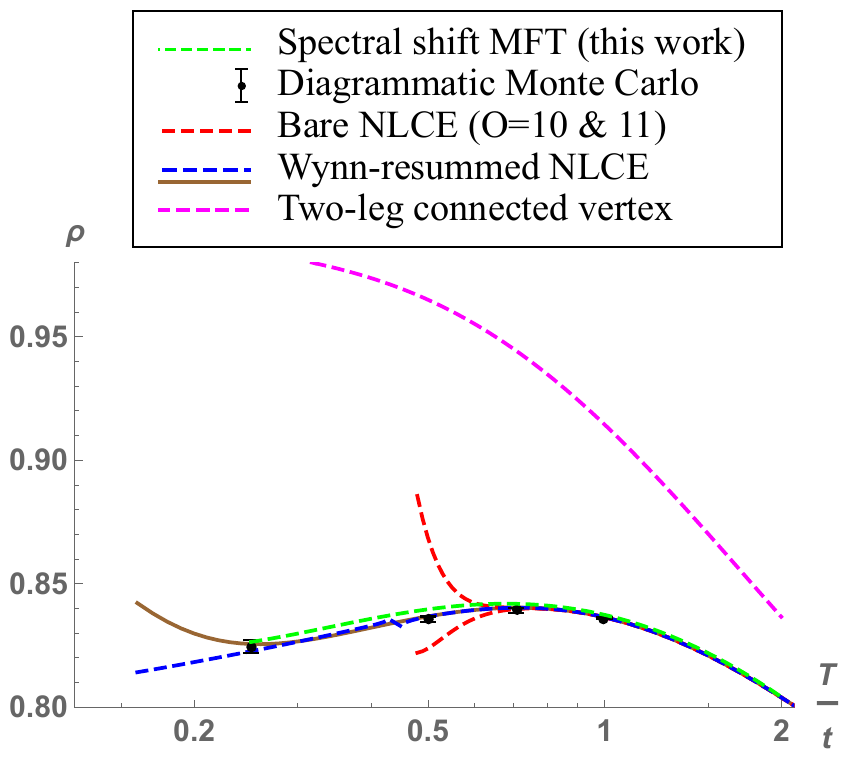}
\caption{
{\bf  Equation of state.
}
The filling factor as a function of temperature obtained in the strong coupling limit ($U=\infty$) with $\mu/t=2$, obtained with several numerical protocols. Controllable results for this observable have previously been computed using numerical linked cluster expansion (NLCE) \cite{PhysRevE.89.063301}: The bare series converges down to $T/t\approx \sqrt{1/2}$, beyond this point, comparison has to be made to Wynn-resummed data. Strong-coupling diagrammatic Mote Carlo is also a controllable technique, with results available down to $T/t=1/4$ \cite{PhysRevB.103.195147}. These show excellent agreement with both bare and Wynn-resummed NLCE data indicating that this regime is indeed under control. 
Spectral shift mean-field theory provides results that are very close to NLCE/Monte Carlo results, but underestimates the carrier density slightly. This reflects the fact that it neglects certain classes of kinetic terms that provide a small but finite correction to the effective bandwidth. To illustrate the impact of the spectral shift, we also include results for the conventional two-legged connected vertex (dashed purple line), which is obtained by solving Eq. (\ref{mft}) using $V$ rather than $\tilde{V}$. This confirms that mean-field-like terms account for almost the entire correction to the carrier density. 
}
\label{result}
\end{figure}

In Fig. \ref{result} we compare the mean-field approach (\ref{mft}) to results obtained from NLCE and SCDMC, which are two controllable techniques that are asymptotically exact in the limit of large clusters and an infinite series expansion, respectively. 
The mean-field results trace these findings closely but do consistently underestimate the carrier density slightly. This is a consequence of the fact that the mean-field theory neglects certain classes of kinetic processes, thus slightly underestimating the carriers' effective bandwidth. 

The importance of mean-field-like processes can be illustrated by also solving Eq. (\ref{mft}) using the unshifted vertex $V$ rather than $\tilde{V}$ (dashed magenta line in Fig. \ref{result}). 
This immediately reveals that almost the entire correction from delocalization effects is contained in the spectral shift. 
Using the shifted vertices as the building blocks of a new spectral-shift expansion, we can implicitly take these corrections into account and expand only in diagram topologies that depend explicitly on inter-site correlations. This does not only provide a starting point that is much closer to the full theory but also reduces computational complexity as the number of diagram topologies is significantly reduced. 

\section{Discussion}
In conclusion, we have derived a framework that integrates mean-field-like kinetic processes from strong-coupling expansions. This results in a shifted expansion conducted around a state with a shifted quasi-particle spectrum where these processes have already been taken into account. Our benchmarks show that mean-field calculations based on this approach--which correspond to order zero in the shifted expansion--contain almost the entire correction to the equation of state. 

A natural comparison for the spectral shift is the summation of self-retracing paths, pioneered by Brinkman and Rice \cite{brink}. The key idea of this method is to approximate the effective bandwidth of a dopant in a Mott-insulator from the set of paths that are independent of the spin background. This provides surprisingly accurate results of the kinetic energy and does also give reliable predictions for the survival probability of the initial state when considering the time-dependent problem \cite{PhysRevLett.116.247202}. 
The method presented here relies on a similar insight: Wide classes of kinetic processes do not chart inter-site correlations. However, while the self-retracing paths represent an approximation, the spectral shift provides a recipe for systematic expansion around the mean-field solution, which is completely unbiased. 

Hence, the spectral shift can radically expand the applicability of strong-coupling diagrammatic Monte Carlo by permitting an expansion purely in diagram topologies that explicitly depend on inter-site correlations. 
Currently, the experimental progress with correlated systems using ultracold atomic gases is very rapid \cite{Gross995,Chiu251,Mazurenko2017,Koepsell2019,koepsell2020microscopic,Cheuk1260,Parsons1253}. 
With the right analytical tools, diagrammatic Monte Carlo can play a crucial role in this development by providing theoretical input with unparalleled precision.

This work was supported by the Swedish Research Council (VR) through grant 2018-03882. Computations were performed on resources provided by the Swedish National Infrastructure for Computing (SNIC) at the National Supercomputer Centre in Linköping, Sweden.
The author would like to thank Marcos Rigol for providing NLCE results for benchmarking. 
\bibliography{biblio.bib}

\end{document}